\documentclass{amsart}
  \usepackage{amssymb}
  \usepackage{amsmath}
  \usepackage{paralist}
  \usepackage{graphics} %% add this and next lines if pictures should be in esp format
  \usepackage{epsfig} %For pictures: screened artwork should be set up with an 85 or 100 line screen

  \usepackage{hyperref} %% Warning: when you first run your tex file, some errors might occur, please just
\theoremstyle{definition}

\usepackage[latin2]{inputenc}
\usepackage[T1]{fontenc}
\usepackage{amsfonts}
\usepackage{mathrsfs}
\usepackage{multicol, graphicx}

\newcommand{\ro}{\,\mathcal{R}_0}

\usepackage{color}

\title[Age specific vaccination scheduling]{Modelling the strategies for age specific vaccination scheduling during influenza pandemic outbreaks}
\author[Di\'{a}na Knipl, Gergely R\"ost]{}

\date{August 31st}

\begin{document}

\maketitle

%% Enter the first author's name and address:
\centerline{\scshape Di\' ana H. Knipl }
\medskip
{\footnotesize
 %% please put the address of the first author
   \centerline{Bolyai Institute, University
of Szeged, Hungary, H-6720 Szeged, Aradi v\'ertan\'uk tere 1.}
   
} %% Do not forget to end the {\footnotesize by the sign }

\medskip

\centerline{\scshape Gergely Röst }
\medskip
{\footnotesize
 %% please put the address of the first author
 \centerline{Analysis and Stochastics Research
Group, Hungarian Academy of Sciences}
   \centerline{Bolyai Institute, University
of Szeged, Hungary, H-6720 Szeged, Aradi v\'ertan\'uk tere 1.}
}

\bigskip

%% The name of the associate editor will be entered by an editorial staff
 \centerline{(Communicated by ....)}

\begin{abstract}
Finding optimal policies to reduce the morbidity and mortality of the ongoing pandemic is a top public health priority. Using a compartmental model with age structure and vaccination status, we examined the effect of age specific scheduling of vaccination during a pandemic influenza outbreak, when there is a race between the vaccination campaign and the dynamics of the pandemic. Our results agree with some recent studies on that age specificity is paramount to vaccination planning. However, little is known about the effectiveness of such control measures when they are applied during the outbreak. Comparing five possible strategies, we found that age specific scheduling can have a huge impact on the outcome of the epidemic. For the best scheme, the attack rates
were up to 10\% lower than for other strategies.  We demonstrate the importance of early start of the vaccination campaign, since ten days delay may increase the attack rate by up to 6$\%$. Taking into account the delay between developing immunity and vaccination is a key factor in evaluating the impact of vaccination campaigns. We provide a general framework which will be
useful for the next pandemic waves as well.

\bigskip


\end{abstract}

\let\thefootnote\relax\footnotetext{$^*$ Corresponding author. Email: rost@math.u-szeged.hu}

%\begin{multicols}{2}
\section{Introduction}

%%---overview of current situation

Developing strategies for mitigating the severity of the influenza epidemics is a top public health priority. As soon as vaccine became available, vaccination campaigns started in several countries as a primary mitigation strategy against the first wave of the 2009 A(H1N1) pandemic. Mathematical models are powerful tools for evaluating intervention strategies and quantifying the potential benefits of different options (Moghadas {\it et al.} 2009). We use a compartmental system with five age groups representing the European population, that incorporates transmission dynamics based on social contact profiles from survey data (Mossong {\it et al.} 2008) and vaccination status. 
Vaccination has not only direct benefit to the individual, but also reduces the risk for those who have not been vaccinated. Giving priority to groups who are the most responsible for spreading the infection can be a benefit to other groups as well. 
%%---previous studies---

US officials have announced that over several months, vaccines for up to 20\% of the population per month could be delivered (Robinson 2009), although later they were facing difficulties with providing supplies. Hungary has a similar maximal production capacity in the range of up to 5\% of the population per week (Falus 2009), thus here we targeted a 60\% vaccine coverage by the end of a three months campaign. Recent studies revealed the importance of age specific intervention strategies, and discussed how to distribute vaccines among age groups, assuming preseasonal vaccination (Longini \& Halloran 2005; Medlock \& Galvani 2009). Here we focus on a completely different aspect: how to prioritize the different groups in the timing of the schedule. Therefore for the purposes of this study, in the baseline scenario we assume that a 60\% total coverage is reached within each age group by the end of the vaccination campaign, and we compare different strategies for the order and the timing of vaccinating different age groups during the campaign. Furthermore, a range of different scenarios are considered in the sensitivity analysis. The relatively low basic reproduction number of influenza (see Chowell {\it et al.} 2007 and Chowell {\it et al.} 2006 for estimations of the reproduction numbers for seasonal and pandemic influenza in temperate countries) implies that effective measures such as vaccination before the pandemic wave could be important, however little is known about the effectiveness of such measures during a pandemic wave. 

%%---novelty of our approach---

As it could be observed all around the world, vaccination campaigns and A(H1N1)v  outbreaks overlapped in time, thus the preseasonal vaccination assumption does not hold. There is an ongoing race between the vaccination campaign and the dynamics of the outbreak, hence it is necessary to implement a dynamic modelling of the interplay of the vaccination and the influenza transmission.

Age structured models are necessary for multiple reasons: various age groups have very different contact profiles thus playing different roles in transmitting the disease. Furthermore, age specific susceptibility, infectiousness, vaccine efficacy and mortality patterns are
also important issues. We developed a compartmental model to track five age groups. We incorporated the fact that it takes about two weeks to develop antibodies and
acquire immunity after vaccination, and during this intermediate period an individual might contract the disease (Nichol 1998). This time delay is shown to be a significant factor during the outbreak. To our knowledge this is the 
first modelling study which reckons with that.

%%---brief summary---

Here we evaluate vaccination strategies for two outcome measures: overall attack rate and total deaths estimated from recent mortality data of the novel A(H1N1)v  pandemic (Donaldson {\it et al.} 2009 ). Though there are several other possible interventions (antiviral treatment, social distancing, school closures etc., see Alexander {\it et al.} 2007, 2009 ; Ferguson {\it et al.} 2006; Merler {\it et al.} 2009; Moghadas {\it et al.} 2008, 2009; Gojovic {\it et al.} 2009)  to mitigate the burden of the outbreak, here we focus only on vaccination. 
The effect of other control measures can be taken into account in a simplified
way by lowering the reproduction number. To show the robustness of our results, a sensitivity analysis was performed with respect to several key model parameters, such as the basic reproduction number, vaccine efficacy, epidemiologic characteristics of the virus, moreover the duration and the intensity of the vaccination campaign.

In the following, we detail the model structure, discuss our results and their epidemiological implications, and place them in the context of the ongoing battle against the nascent A(H1N1)v  pandemic. Our aim was to develop a model that can be applied to explore the effect of alternative vaccination strategies, and this model can also be used for predicting the future incidence of cases in the next wave of pandemic influenza.

\section{The model}

Our model is a compartmental differential system, based on the classical SEIR ($S$usceptible, $E$xposed, $I$nfective, $R$ecovered) model.
We have incorporated three features to develop a more realistic model: 

i) we introduced age structure with five age groups, where the contacts between age groups are derived from the European survey Mossong {\it et al.} 2008,

ii) we account for the
fact that it takes up to 14 days after vaccination to produce sufficient amount of antibodies to provide immunity (CDC 2009, Nichol 1998),

iii) we did not assume vaccination before the outbreak, because we want to model the interplay between the dynamics of the epidemics and the vaccination campaign. 

It is assumed that the transmission of infection occurs through close contacts between susceptibles and infected individuals, where for simplicity, mass action incidence is used (Alexander {\it et al.} 2008, 2009; Arino {\it et al.} 2006). Exposed individuals in the class $E$ cannot transmit the disease in the latent period, during which viral titres increase to detectable and transmissible levels (Baccam {\it et al.} 2006). Since the latent period is relatively short, we neglected the small probability of someone receiving the vaccine while being in the class $E$. We assume further that infected and recovered individuals will not be vaccinated, therefore vaccination is administered only for individuals in the $S$ class until we reach the targeted 60\% coverage on the population level.
If the pool of susceptibles depleted before reaching that coverage, we stopped vaccination in the simulations. However, contacts of cases may have already been infected by the time that vaccine is taken up, because high proportion of infection is not laboratory confirmed, and also asymptomatic or atypical infections can also occur. The magnitude of such limitation is less of a problem when considering the first epidemic wave of pandemic influenza than for the second or third wave. The model tracks 5 age groups, distributed as the 2005 European Union population (Eurostat 2006). We considered the case when a single dose is sufficient to induce immunity. A single dose is recommended for the Hungarian vaccine (Johansen {\it et al.} 2009; Vajo {\it et al.} 2009), and according to the most up-to-date recommendation of the European Medicines Agency that can be sufficient for other vaccines as well, see also (Nishiura \& Iwata 2009; Greenberg {\it et al.} 2009). The five age groups we considered are 0-9, 10-19, 20-39, 40-64 and 65+ years old. In the model equations we use the upper index $i=1,...,5$ to denote the corresponding age group, respectively. Based on the most recent serological data (Miller {\it et al.} 2010), we assumed no pre-existing immunity in the first four age groups,
and 20$\%$ reduction in susceptibility in the elder age group.

Vaccinated individuals move into the class $W$ for an intermediate period,
during that infection is still possible. After 14 days they become immune with probability $q$, and move into the $R_W$ class, or if the vaccine was ineffective, then they move into $S_V$, meaning that they are still susceptible to the disease despite having been vaccinated. Such individuals will not receive the vaccine again, but can come into contact with infectious individuals and contract the disease. It is assumed in the baseline scenario that the same
epidemiological parameters apply to these individuals as to the non-vaccinated, i.e. $\mu_{E_V}=\mu_E$ and $\mu_{I_V}=\mu_I$, except that we assumed a reduction of infectiousness for unsuccessfully vaccinated persons (see Table 1). This factor $\delta$ is set to be 0.75 in the baseline scenario, but other possibilities for the parameters of unsuccessfully vaccinated individuals are also discussed in the sensitivity analysis. The epidemiological parameters, such as the length of incubation and infectious periods, vaccine efficacy, transmission rates etc. are taken from the literature and discussed in detail in Section 3.

The transmission diagram of our model is presented in Figure 1, without the
age structure.

Thus, altogether we have 50 compartments and the following non-autonomous system of 50 differential equations

\begin{equation}\aligned
\dot S^{i}&=-S^{i}\lambda^i-V^{i} & \dot S_{V}^{i}&=(1-q_{i})\mu_{W} W^{i}-S_{V}^{i}\lambda^i
\\
\dot E^{i}&=(S^{i}+W^{i}) \lambda^i-\mu_{E}^{i}E^{i} &\dot E_{V}^{i}&=S_{V}^{i} \lambda^i-\mu_{E_{V}}^{i}E_{V}^{i}
\\
\dot I^{i}&=\mu_{E}^{i}E^{i}-\mu_{I}^{i}I^{i} & \dot I_{V}^{i}&=\mu_{E_{V}}^{i} E_{V}^{i}-\mu_{I_{V}}^{i} I_{V}^{i}
\\
\dot R^{i}&=\mu_{I}^{i}I^{i} & \dot R_{V}^{i}&=\mu_{I_{V}}^{i}I_{V}^{i}
\\ \dot W^{i}&=V^{i}-W^{i} \lambda^i - \mu_W W^{i}  & \dot R^{i}_{W}&=q_{i} \mu_{W} W^{i}
,\nonumber\endaligned\end{equation}

where the force of infection is given by $$\lambda^i=\sum_{j=1}^5 (\beta_{j,i}(I^{j}+\delta I_{V}^{j})) $$ and the upper index
$i=1,..,5$ represents the corresponding age group. Here $V^{i}=V^{i}(t)$ are the prescribed piecewise constant vaccination rate functions determined by the specific strategy.
In our equations we ignored mortality, since even a 40\% attack 
rate and 0.05\% case fatality rate cause very minor changes in the demographic
scale, and the number of disease induced deaths can be derived simply from the
attack rate and the case fatality ratio. Nevertheless, the age specific mortality is an important issue, and such information can be obtained from the age specific attack rates calculated by the model and the mortality patterns when they are available, see Section 4 and 5. We start the model at $t=0$, with time measured in days, introducing a small number of infectives into the population. The time $T$ refers to the delay in start of the vaccination campaign, thus $V^i(t)=0$ for any $t<T$, and in a case of a three months campaign, $V^i(t)=0$ for $t>T+90$ as well.

\begin{figure}
\centerline{\includegraphics[height=4.5cm]{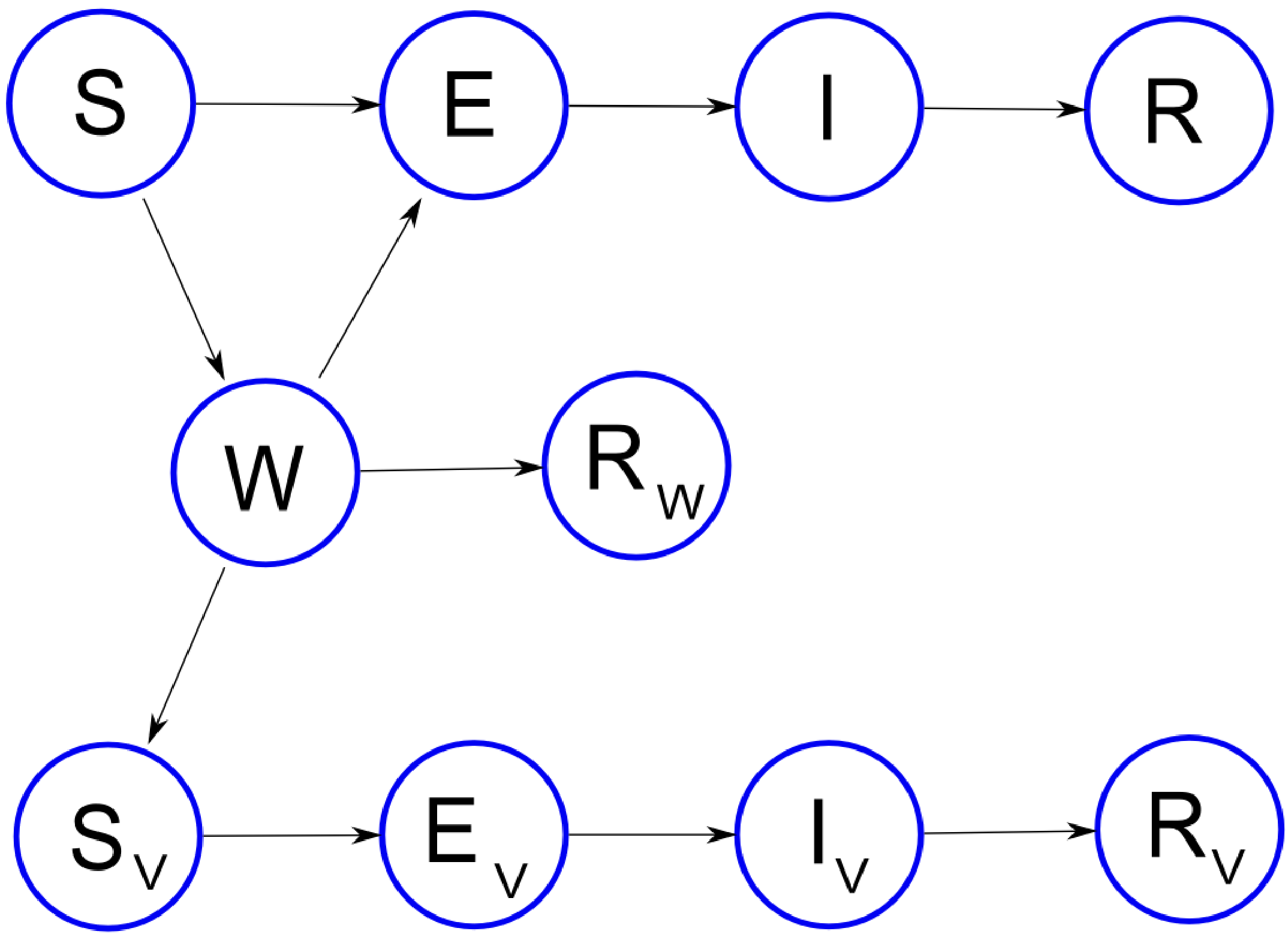}}
\caption{Transmission diagram without age structure. Here $W$ is the
compartment of those who have been recently vaccinated, $R_W$ who have already acquired immunity after vaccination, and the subscript $V$ denotes the classes
of unsuccessfully vaccinated
people.}
\end{figure}

\section{Parameters}
The epidemiological parameters are summarized in Table 1. So far there are no precise clinical estimates of the basic model parameters $\mu_E$ and $\mu_I$ defining the inverse average exposed and infectious time durations, hence we used plausible values from the range estimated in Balcan {\it et al.} 2009 . We used the age distribution of the European Union (Eurostat 2006) for our simulations. 
For vaccine efficacy we used the same parameters as in Medlock \& Galvani 2009, obtained from Basta {\it et al.} 2008.
In the sensitivity analysis we considered various values for the reduction of infectiousness $\delta$ and vaccine efficacy against infection $q_i$. The value of $\mu_W$, representing the inverse time duration needed to develop
antibodies to gain immunity, is the same as  US CDC 2009 uses, being in accordance with Nichol 1998.

\begin{table}[t]
\centering
\begin{tabular}{c c c c}
\hline
parameter & notation & value (range) & source \\ 
\hline
latent period & $1/\mu_E$ & 1.25 days & Balcan {\it et al.} 2009 \\
infectious period & $1/\mu_I$ & 3 days & Balcan {\it et al.} 2009 \\
vaccine efficacy for 0-65 yrs old & $q_i, i=1..4$ & 0.8 (0.7-0.9) & Basta {\it et al.} 2008 \\
vaccine efficacy for 65+ yrs old & $q_5$ & 0.6 (0.5-0.7) &  Basta {\it et al.} 2008 \\
transmission rates & $\beta_{i,j}$ & see 3.2 & Mossong {\it et al.} 2008 \\
time to develop antibodies & $1/\mu_W$ & 14 days & Nichol 1998\\
reduction in infectiousness & $\delta$ & 0.75 (0.5-1) & \\
vaccination coverage &  & 60\%(30\%-80\%) & \\
basic reproduction number & $\ro$ & 1.4 (1.2-1.8) &  Truite {\it et al.} 2009\\
\hline
\end{tabular}
\caption{Description of the model parameters}
\label{tab:}
\end{table}

\subsection{Baseline scenario}

The severity of the influenza outbreak and the initial rate of increase depends on the basic reproduction number $\ro$, which is among the most urgently estimated quantities in a pandemic situation, and defined as the average number of secondary infectious cases caused by an infected individual in a population consisting of susceptibles only (Diekmann {\it et al.} 2010; Heffernan {\it et al.} 2005).
In the baseline scenario we assumed that the basic reproduction number $\ro
$ is 1.4. This value seems to be reasonable (Truite {\it et al.} 2009) and similar to mean estimates of seasonal infuenza in temperate countries , however, we considered milder and more severe cases ($\ro=1.2$, $\ro=1.6$, $\ro=1.8$) as well. We run computer simulations for a population of $N=100 000$. Due to the special form of the equations, the results are scalable and the attack rates (defined as the fraction of susceptibles who do not contract the disease during the course of the outbreak) remain the same for populations of any size. In the baseline scenario, after the start of the vaccination campaign at time $T$, vaccine is administered to 0.667\% of the population per day, thus reaching a coverage of 60\% at the end of the campaign at time $T+90$. This is an intensive, but realizable vaccination plan (Falus 2009). The distribution of these vaccines among the age groups on any given day is determined by the specific strategy (see Section 4).
Based on European Union statistics Eurostat 2006, we use the following age distribution per 100 000 citizens:
$$\begin{array}{|c|c|c|c|c|}
\hline  N^1	& N^2 &	N^3	& N^4 & 	N^5 \\ \hline
10500	& 12000& 28500	& 32500 &	16500 \\ \hline
\end{array}\quad ,$$
where $N^i$ is the total number of individuals in age group $i$, and $N=\sum_{i=1}^5 N^i.$

\subsection{The contact matrix and the basic reproduction numbers}

There are practical reasons for using five age groups besides for simplicity: we think that targeting smaller groups is not feasible in a real situation, and we can not expect to have vaccination data with such details. Also, in view of the 
data in Mossong {\it et al.} 2008, the population can be divided into these five age groups with more or less similar contact profiles within these larger groups, which are then easier to be targeted. 
Following Medlock \& Galvani 2009, we used the European survey data Mossong {\it et al.} 2008 to estimate
the transmission rates between age groups. That survey dealt with 17 age groups
(0,1-4,5-9,10-14,..,70-74 and 75+) in 8 European countries. Since we work with different age groups, after averaging the European contact matrices from the 8 countries weighted by their population size, we calculated a $5\times 5$ contact matrix $\bar C =(\bar c_{i,j})$, using weighted averaging based on EU age distribution (Eurostat 2006) to combine smaller age groups into the larger ones. Since contacts are assumed to be mutual, following Medlock \& Galvani 2009, we performed a symmetrization procedure to ensure that the total number of contacts between
two age groups are consistent: we set $$c_{i,j}=\frac {\bar c_{i,j} N^i+\bar c_{j,i} N^j}{2N^i} .$$
Eventually, we obtained the contact matrix

$$\mathcal{C}=  \left( \begin{array}{ccccc}
 5,3580	& 1,0865 &	3,0404	& 2,4847 & 	0,8150 \\
0,9507	& 10,2827& 2,8148	& 3,6215 &	0,7752 \\
1,1201	& 1,1852 &	6,5220	& 4,1938	& 0,9016 \\
0,8027	& 1,3372 &	3,6776	& 5,2632 &	1,3977 \\
0,5187	& 0,5638 &	1,5573	& 2,7531&	2,0742
\end{array} \right), $$
where the elements $c_{i,j}$ represent the number of contacts an individual in age group $i$ has with individuals in age group $j$, and satisfy $c_{i,j}N^i=c_{j,i}N^j.$

Age specific contact rates can be converted to the age specific transmission rates $\beta_{i,j}$ as follows. An infected individual in age group $i$ has $c_{i,j}$ contacts with individuals in age group $j$, at some time $t$ a fraction $S^j(t)/N^j$ of those contacts are with susceptibles, $S^j_V(t)/N^j$ with vaccinated susceptibles and $W^j(t)/N^j$ with recent vaccinees. Thus, we obtain that the rates  of infections in age group $j$ by individuals in age group $i$ is given by  
$$\beta c_{i,j} \frac{S^j(t)}{N^j} I^i(t), \quad \beta c_{i,j} \frac{\delta S^j_V(t)}{N^j} I^i(t), \quad \beta c_{i,j} \frac{W^j(t)}{N^j} I^i(t),$$ where $\beta$ is a transmission parameter, which involves the normalization of the contacts to unit time and the infectiousness of the virus. By defining $\beta_{i,j}=\beta c_{i,j}/{N^j}$, we obtain the model equations and force of infections described in Section 2. 
Using the transmission rates we can construct the next generation matrix $\mathcal{N}$ (see Diekmann {\it et al.} 2010), and then fit $\beta$ numerically to ensure that the basic reproduction number, which is the dominant eigenvalue of the next generation matrix, is equal to 1.4. 

Since in the early phase of the pandemic $S^{j}_{0}\approx \varrho^j N^j$ (where $\varrho^j=1$ for $j=1,2,3,4$ and $\varrho^5=0.8$), the number of infections generated in age group $j$ by an infected individual from age group $j$ is given by ${\beta_{i,j} S^{j}_{0} }/{\mu_{I}}$. The elements of the next generation matrix are given by this formula, i.e. 
$$\mathcal{N}=(n_{i,j})=\Big(\frac{\beta_{i,j} S^{j}_{0} }{\mu_{I}}\Big)_{i,j}=\Big(\frac{\beta \varrho^j c_{i,j}}{\mu_{I}}\Big)_{i,j} \quad,$$
thus they can be obtained from the elements of the contact matrix.

By standard numerical procedure we find that if $\beta=0.0334$, then the dominant eigenvalue of $\mathcal{N}$, which is the basic reproduction number
$R_{0}$, equals to 1.4, thus in the baseline scenario we use this $\beta$ value. 

Note that in this scenario $n_{2,2} >1$, which means that there is a self sustaining outbreak in the age group of 10-19 years old individuals.

\section{Strategies, simulations and results}

We simulate, evaluate and compare five possible strategies for the vaccination
schedule. In each of the strategies, vaccine is administered to about 0.667$\%$ of the population daily and a 60$\%$ vaccination coverage is achieved by the end of the campaign in every age groups. Each strategy determines the piecewise constant vaccination functions $V^i(t)$ straightforward. The overall attack
rates and the final size of the epidemic are estimated from taking the values
of the non-infected classes after the pandemic wave, at $t=250$.

The case fatality ratio is yet to be determined for many countries, however estimated rates are coming down for industrialised countries to as low as 0.02\% (ECDC 2009, Presanis {\it et al.} 2009; meaning that one in five thousand cases has a fatal outcome). Most recent risk assessments suggest that in European countries the overall fatality rate may be less or similar to a moderate influenza season. However, age specific rates are expected to show a very different picture, with higher mortality in younger age-groups. Here we used
the recently published data (Donaldson {\it et al.} 2009, Vaillant {\it et. al} 2009), and calculated with  the mortality rate {20:12:30:60:80} fatal outcomes per 100 000 cases in the five age groups to compare the five strategies, but other patterns are also discussed in the sensitivity analysis. 

\subsection{The five strategies} Here we describe the five strategies we compare. For each strategy, 60\% overall vaccination coverage is reached by the
end of the campaign in the baseline scenario.

\medskip

{\bf A - Conventional strategy} 

\medskip

This strategy consists of three phases and we constructed it to resemble the official strategy of Hungarian health authorities. Similar strategies have been proposed for many other countries.

\textit{Phase 1 :} 42 days, vaccination of high risk groups, elder people, emergency and health care personnel, workers of critical infrastructure facilities. According to the age specific breakdown (which is described in Section 3.1), 
2000, 16000, 10000 people are being vaccinated in age groups 0-19, 20-64, 65+, respectively. 

\textit{Phase 2 :} Vaccination campaign in schools for 18 days, 12000 children of
age $\leq$ 19 years old.

\textit{Phase 3 :} In the last 30 days, vaccination is given to the general population such that we achieve the 60$\%$ coverage in each group by the
end of this phase. 

\medskip

{\bf B - Uniform strategy} 

\medskip

This is the universal vaccination strategy, when there are no prioritized age groups, so we assume that vaccination is completely random and 0.667$\%$ of each age group is vaccinated daily, throughout the 90 days.

\medskip

{\bf C - Elderly first strategy} 

\medskip

Phased vaccination of elder people (age $\geq$ 65 years old) first up to 60 $\%$ coverage before vaccine is delivered to the other part
of the population (age $\leq$ 64 years). This two-phase strategy is similar to
plans usually implemented against seasonal flu.

\textit{Phase 1 :} 15 days, 65+ years old

\textit{Phase 2 :} 75 days, 0-64 years old

\medskip

{\bf D - Children first strategy} 

\medskip

Phased vaccination of children (age $\leq$ 18 years old)
first up to 70 $\%$ coverage before vaccine is delivered to adults was studied
in Yang {\it et al.} 2009. The only difference here in the D strategy is that coverage goes up to only 60$\%$, and our prioritized age group is the 0-19 years old. Specifically:

\textit{Phase 1 :} 20 days, 0-19 years old

\textit{Phase 2 :} 70 days 20+ years old

\medskip

{\bf E - By contacts strategy}

\medskip
 
Here we take advantage of the full contact structure of our five age groups, and vaccinate them in five phases, according to the decreasing order of their total contact numbers.

\textit{Phase 1 :} 10-19 years old, 11 days

\textit{Phase 2 :} 20-39 years old, 26 days

\textit{Phase 3 :} 0-9 years old, 10 days

\textit{Phase 4 :} 40-64 years old, 29 days

\textit{Phase 5 :} 65+ years old, 15 days

\begin{figure}[tbh]
\centerline{
\includegraphics[height=7cm]{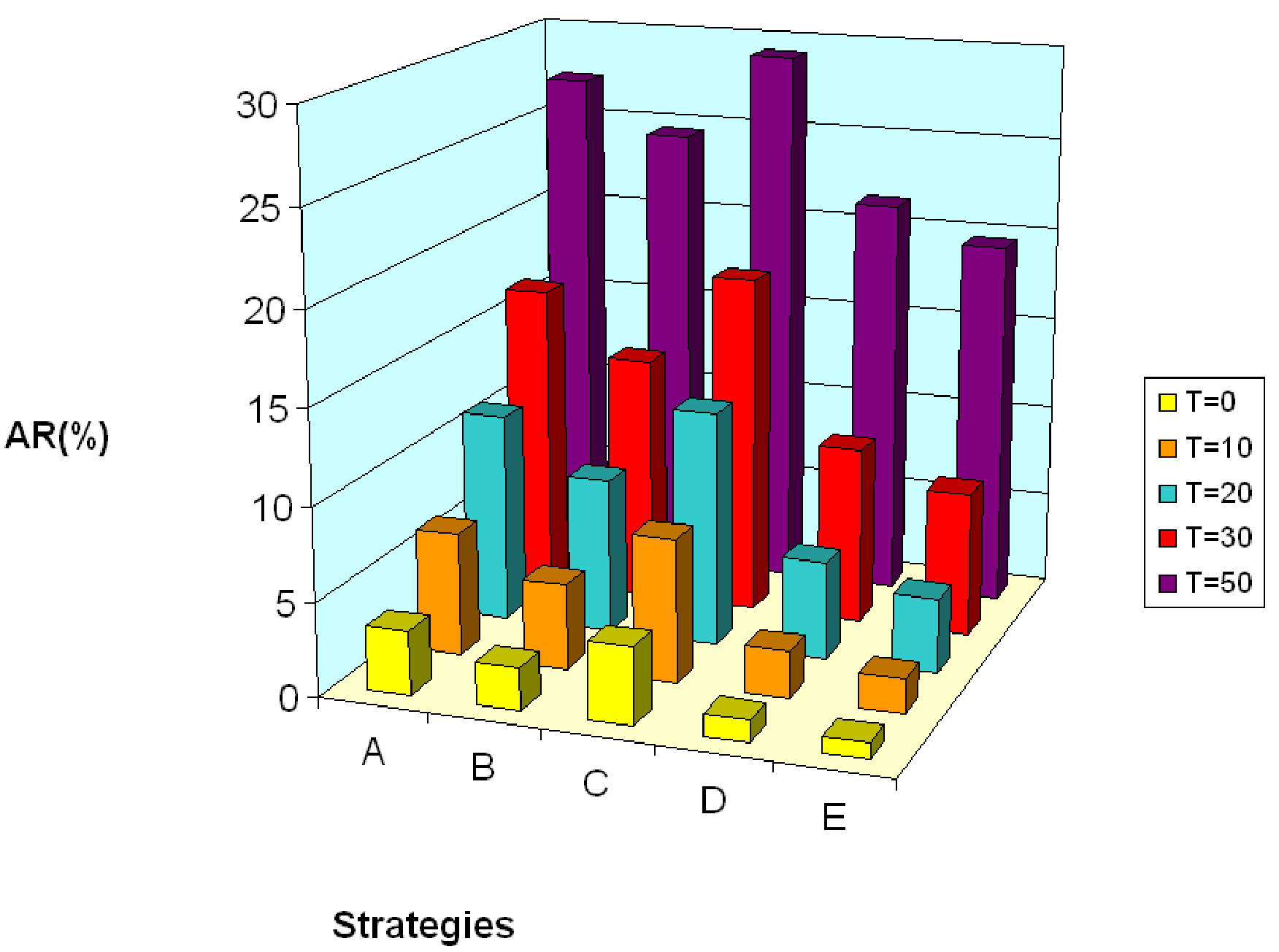}
}
\caption{Overall attack rates of the five strategies for various delays $T$ in starting the vaccination campaign (A - Conventional, B - Uniform, C - Elderly
first, D - Children first, E - By contacts) for $\ro=1.4$. Without vaccination, the attack rate is 42\%. }
\end{figure}

\subsection{Main results}

We have evaluated and compared the above described five strategies for various delays in start of the vaccination. Increasing population level immunity may slow down the spread of infection, reduce the height of the epidemic peak thus decreasing the pressure on health care facilities, and significantly reduce
morbidity and mortality of pandemic infections. Thus our main outcome measures are the age specific attack rates and number of fatal cases. 

The overall attack rates are summarized in Figure 2 and Figure 3. We can see that by means of illness attack rate, the strategy $E$ gives the best result, followed by $D,B,A$ and $C$ in that order, whenever we start the vaccination campaign.
In case of an early start on day 1, all of the five strategies are effective
with attack rates between 1-4 $\%$. As the delay in start of vaccination increases, the significance of our choice of strategy becomes more apparent.
We can advert the infection of 10 $\%$ of the population by choosing strategy $E$ instead of $C$ if the campaign starts between days 30 and 45. If vaccination
starts on day 50, the attack rates will be between 21-29 $\%$. For the sake of
comparison we note that in the absence
of a vaccination campaign, $\ro=1.4$ results in an a 42\% attack rate.

The delay $T$ in start of vaccination is a crucial factor on the final size of the pandemics: even for the best strategy, ten days delay can increase the
overall attack rate by up to 6$\%$ (see Figure 3).

\begin{figure}[tbh]
\centerline{\includegraphics[height=6cm]{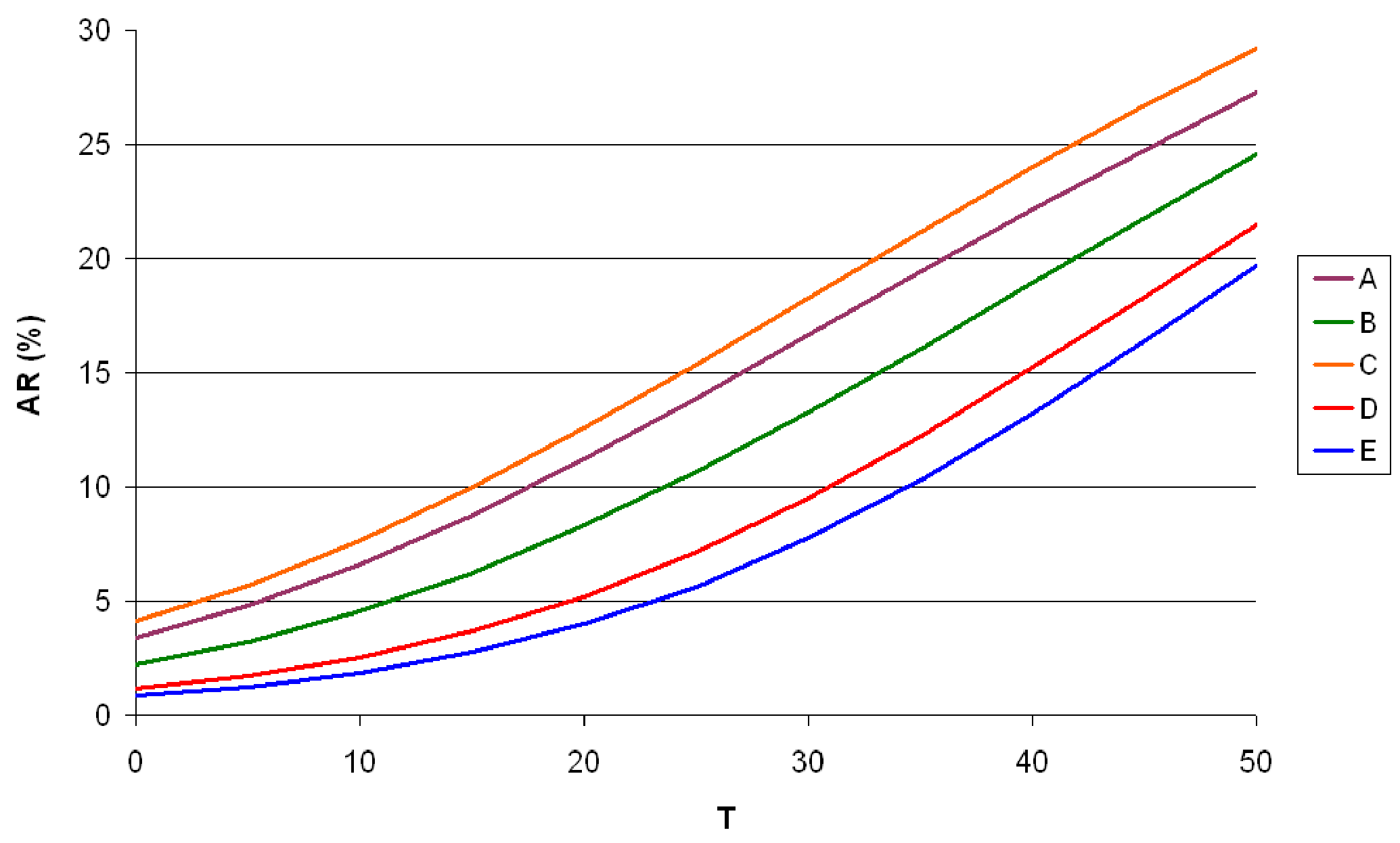}}
\caption{Overall illness attack rates of the five strategies plotted versus the delay $T$ (A - Conventional, B - Uniform, C - Elderly
first, D - Children first, E - By contacts) for $\ro=1.4$. Without vaccination, the attack rate is 42\%. }
\end{figure}

%------age specific attack rates

Figure 4 shows the attack rates in the five age groups for the five strategies and also for the case when there is no vaccination at all.
Depending on which age group we want to protect, we can
choose different strategies. Since age groups peak at different times, 
the delay in start of vaccination may also be a factor in choosing our strategy: for example for $T=50$ the strategy $C$, which gives priority to the elder age group, protects them essentially to the same extent as any other strategy (attack rates slightly over 10\%), while there are large differences in 
the attack rates for the age groups 0-9 and 10-19 years old. For $T=0$ and $T=25$ the strategy $E$ protects the elder better with 0.5$\%$ and 3.2$\%$ attack rates versus the 1.4$\%$ and 5.4$\%$ attack rates that correspond to the strategy $C$ among the elder age group, despite the 
fact that in the strategy $E$ they are the last to get the vaccine. This very interesting result may seem counterintuitive, however some similar findings have already been obtained for preseasonal vaccination in Medlock \& Galvani 2009. Furthermore, the public health implication of this phenomenon is that if the actual policy is to protect a specific age group, then the best strategy depends also on the delay in start of vaccination, therefore authorities must be able to adapt their strategy as the pandemic evolves.
Very high attack rates can be observed among teenagers for strategies $A,B$ and $C$.
 
\begin{figure}[tbh]
\centerline{
  \includegraphics[width=14cm]{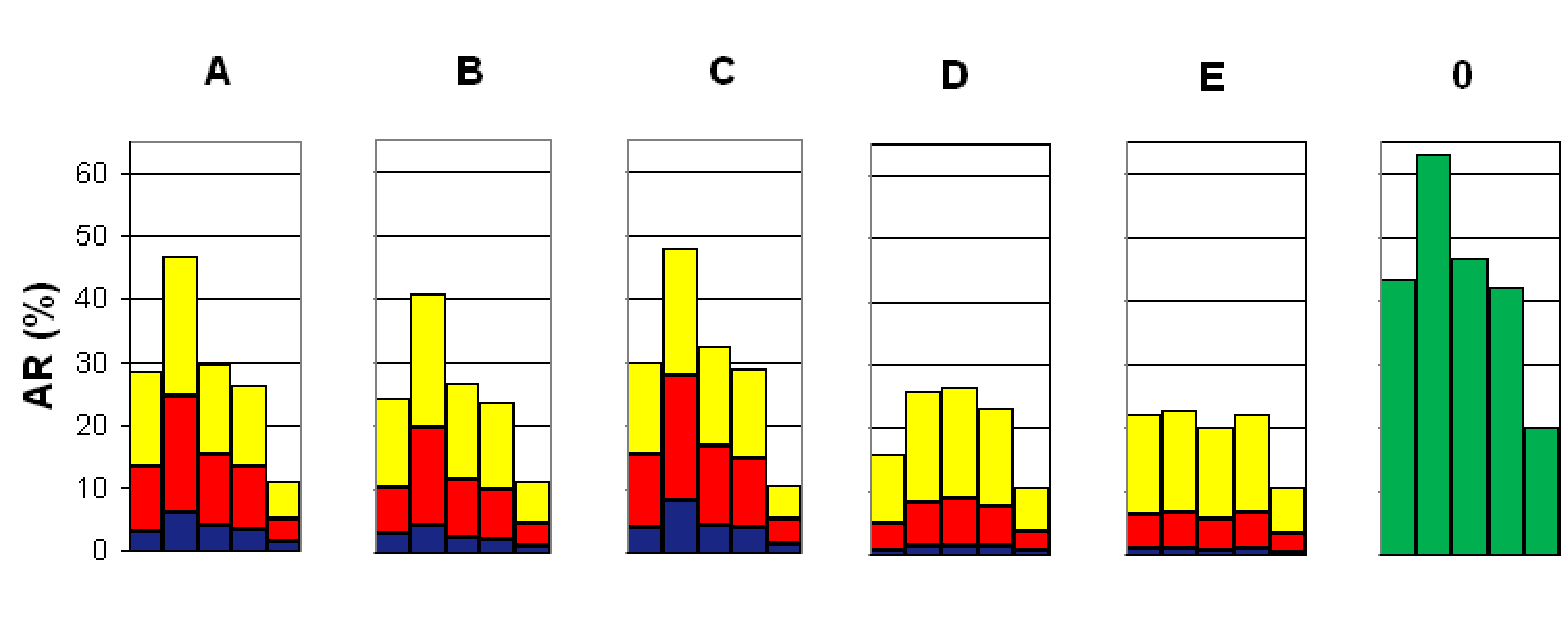}
}
\caption{Age specific attack rates of the five strategies in the age groups 0-9, 10-19, 20-39, 40-64 and 65+ years old, represented by the five
columns, respectively, for various delays $T$ in start of the campaign: $T=0$ (blue), $T=25$ (red) and $T=50$ (yellow) for $\ro=1.4$. The colours indicate the increases in the attack rates for longer delays in start of the campaign. The strategies: A - Conventional, B - Uniform, C - Elderly
first, D - Children first, E - By contacts, 0 - No vaccination.}
\end{figure}
%------mortality E>B>A>D>C

The comparison of the strategies by means of mortality is depicted in Figure 5.
The relation between the strategies  for this outcome measure is similar to the attack rates: strategy $E$ being the best and $C$ the worst. However, we can observe a much bigger difference between the best and the worst schedules: the overall mortality is 50$\%$ higher for the strategy $C$ than for the strategy $E$ for a late
start in vaccination, and six times higher for an early start. The reason of this huge difference is partly that compared to the elderly, the recent H1N1 causes unusually high mortality among children and young adults, who has the most contacts as well and prioritized by strategy $E$. The age specific
details can be seen in Figure 6.

The epidemic curves in the five age groups are plotted in Figure 7 for strategies $A$ and $E$ with $T=25$. Notice that age groups peak in different times, and for the strategy $A$ the teenagers are affected disproportionally by the epidemics.

\begin{figure}
\centerline{
  \includegraphics[width=12cm]{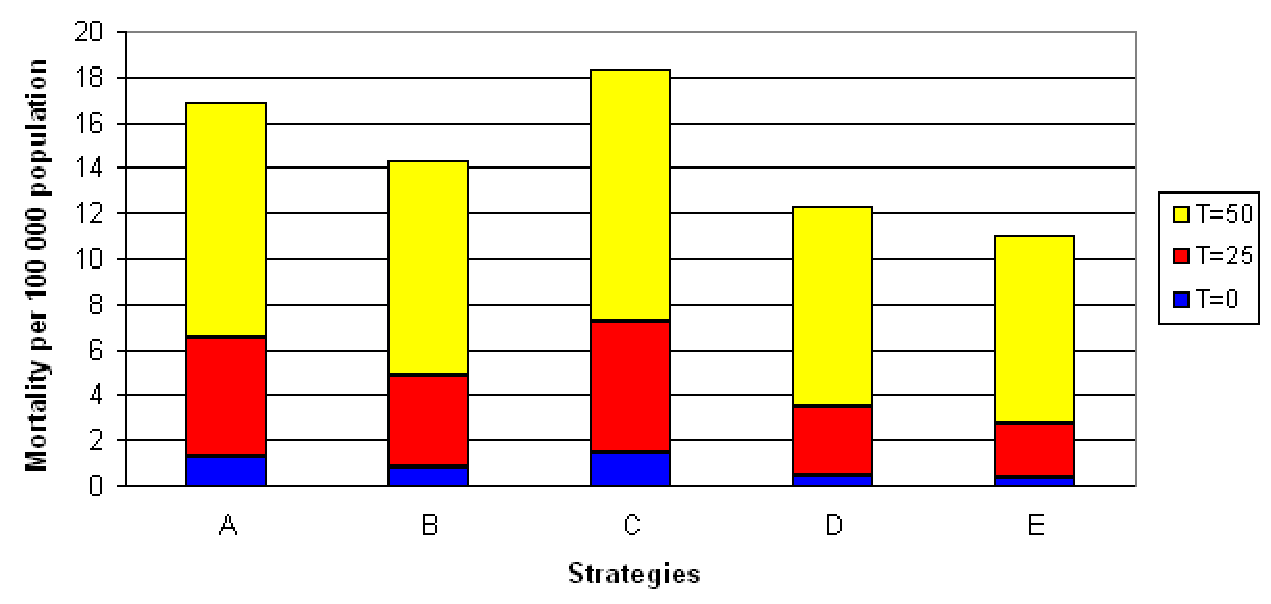}
}
\caption{Overall mortality (number of deaths in a population of 100000)  of the five strategies (A - Conventional, B - Uniform, C - Elderly
first, D - Children first, E - By contacts) for $T=0$, $T=25$ and $T=50$ days
delay in start of the vaccination campaign, using the mortality pattern of the baseline scenario (sum of the products of the age specific attack rates, case fatality rates and population sizes), for $\ro=1.4$. The colours indicate the increases in the attack rates for longer delays in start of the campaign.}
\end{figure}

\section{Sensitivity analysis}
We performed a systematic analysis to reveal the sensitivity to
several key parameters. Our results turned out to be very robust
in the sense that modifying some parameters do not change which
strategy is the best if our outcome measure is the overall attack rate.
However, the mortality pattern is important to select our strategy to minimize the number of fatal cases. Changing several parameters at the same time, we
did not observe any unexpected behaviour.

\medskip

{\bf Basic reproduction number}

\medskip

We consider a less ($R_{0}=1.2$) and a more severe ($R_{0}=1.6$) situation, Figure 7 shows the dependence of the attack rates on $T$.
In the milder case, we can observe that for the early start of vaccination for all five strategies the attack rates are below 1$\%$, the difference between the strategies shows only for delayed start, attack rates being between 2-4$\%$.
Thus a 60$\%$ vaccination coverage is able to prevent a large outbreak even if the campaign starts relatively late. Note that without vaccination, the attack
rate is 22.5$\%$.
In the more severe situation in the case of late start each strategy can mitigate the pandemic only a little and there is not much difference between the attack rates: about 50$\%$ for all strategies. In the absence of vaccination, the attack rate is 55$\%$, thus in this situation the vaccination campaigns are
not effective. There is a significant difference between the strategies if we start the campaign early, in this case with strategy $E$ we have only 5$\%$ attack rate, while for the worst strategy $C$ the 
attack rate is as high as 19$\%$ if $T=0$. 
In the even more severe case $R_{0}=1.8$, for the early start the attack rates
vary between 18 and 36$\%$, strategy $E$ being far the best. However, for the
late start the difference between the strategies mostly vanishes and the attack
rates exceed 60$\%$, being very close to the 64$\%$ attack rate of the no vaccination case.

%\begin{figure}
%\centerline{
%\includegraphics[width=7cm]{epicurve1}\includegraphics[width=7cm]{epicurve2}
%}
%\caption{SOME Epidemic curves / WHICH ONES?? A,E strat T=0,25,50 + 5 korcsoport %A,E T=10?}
%\end{figure}

\medskip

{\bf Vaccine efficacy}

\medskip

Monitoring the serological responses to influenza vaccines alone is not sufficient to establish evidence that a vaccine is effective, because there has been no clear relationship shown between serological response to influenza vaccination and subsequent morbidity and mortality. Vaccine effectiveness measures with clinical outcomes may provide better estimate for the evaluation of how much protection the vaccine may provide, however such estimates are not readily available at this stage of the vaccination campaign against pandemic influenza. Considering less effective ($q_i=0.7$ for $i=1,...4$ and $q_5=0.5$), or more effective vaccine ($q_i=0.9$ for $i=1,...4$ and $q_5=0.7$) against infection (see Basta {\it et al.} 2008, CDC 2008) resulted in slightly lower and higher attack rates, respectively, typically by 0.5-1.5$\%$.

\medskip

\begin{figure}
\centerline{
\includegraphics[width=14cm]{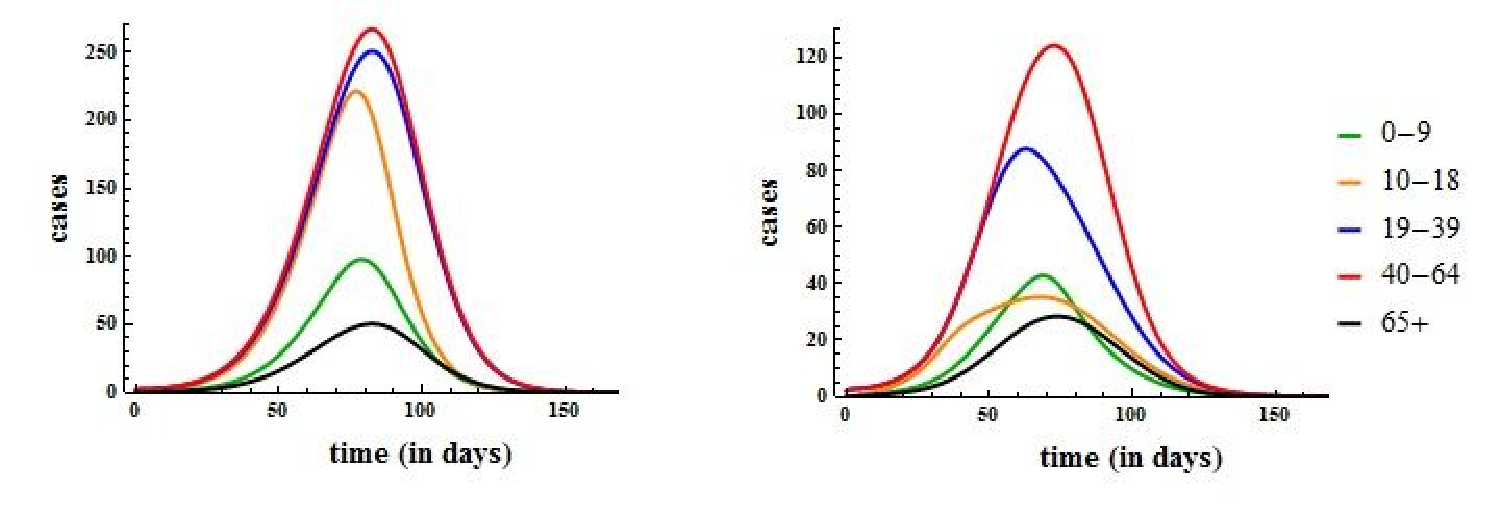}
}
\caption{Epidemic curves (absolute numbers of infections) in the age groups for strategies $A$ (conventional) and $E$ (by contacts), when vaccination starts on day 25, in a population of 100 000 in the baseline scenario ($\ro=1.4$).}
\end{figure}

{\bf Reduction of infectiousness}

\medskip

Different scenarios were considered regarding the infectiousness of vaccinees. Taking $\delta=1$ (no reduction in infectiousness) or $\delta=0.5$ (more significant reduction in infectiousness) did not change much in the outcomes,
the differences in the attack rates were less than 1$\%$, compared to the baseline scenario $\delta=0.75$. Assuming that unsuccessful vaccination shortens
the infectious period by one day in case of infection, i.e. $1/\mu_{I_V}=2$,  the attack rates are decreased by less than 1$\%$.

\medskip

{\bf Intensity of vaccination campaign}

\medskip

For the purposes of this study we assumed that 60\% coverage can be reached
in 90 days in the baseline scenario. In the sensitivity analysis, we compared the strategies $A$ and $E$ with respect to the intensity and the length
of the vaccination campaign, adjusting the phases accordingly. Assuming that we complete the vaccination of 60$\%$ of the population in 120 days (i.e. vaccine is administered to 0.5$\%$ of the 
population daily), then the attack rates  were at most 4$\%$ higher for the
strategy $E$, and 6$\%$ higher for the strategy $A$, depending on the delay in
start of the campaign. A less intensive campaign enlarged the differences between the strategies, which implies that choosing the best strategy is even more important if the vaccine availability  or the capacity of the health care system is limited.

We found another counterintuitive result regarding the strategy $A$: vaccinating
60$\%$ in 120 days gives a worse outcome than vaccinating 45$\%$ in 90 days. The 
reason behind this phenomenon is that if we adjust our phases to the given time intervals, than in the first case the phases will be longer and the most important age groups in the second phase receive the vaccine in such a delay that outweighs the benefits of vaccinating more people, and vaccinations after day 90 has very limited effect on the outcome.

A much less intensive campaign, vaccination of only 30$\%$ in 90 days resulted in drastically higher overall attack rates, up to 12$\%$, depending on the 
strategy and the delay in start of the campaign. Strategy $E$ is consistently
the best option for a less intensive campaign, having 5-6$\%$ smaller attack
rate than strategy $A$ for both early and late start of the campaign.

For a very intensive campaign, when 80\% coverage is reached within 90 days,
the attack rates are lower, for example for strategy $E$ it is 17\% for
$T=50$, 4\% for $T=25$ and below 2\% for $T \leq 16$.

\medskip

{\bf Mortality patterns}

\medskip

For the sake of comparison, we examined the outcomes of the strategies for a mortality pattern which is somewhat different from what we see for A(H1N1)v, namely the 1957 pandemic. Suggested by Serfling {\it et al.} 1967, we assumed that the fatality rates for the five age groups per 100 000 cases have the proportions 0.1:0:0.1:1:4, meaning that mortality is four times higher in the elderly, than in the age group 40-64 years old etc. We observed that
in this case there is no significant difference in the mortalities between
the five strategies, since the vast majority of fatal cases occurs in the elder age groups, where the five strategies resulted in similar attack rates.

It is important to notice that in Figure 5 we overestimated the number of fatal cases for strategy $A$, since then the high risk groups are vaccinated first in
each age group thus proportionally less high risk people will be infected than
in the other strategies, reducing the mortality. This factor of reduction is
hard to determine (see Fleming  \& Elliott 2006 ), since we need to estimate the fraction of high risk individuals in each age group, and also their relative risk for a fatal outcome.
Nevertheless, we can see that the ratio of the attack rates for strategy $E$ and strategy $A$ is 0.3 for the case of an early start, and 0.65 for a late start, thus to fully compensate the
higher attack rate, prioritizing high risk groups in strategy $A$ must result in a significantly lower mortality.

\medskip

\begin{figure}
\centerline{
 \includegraphics[width=14cm]{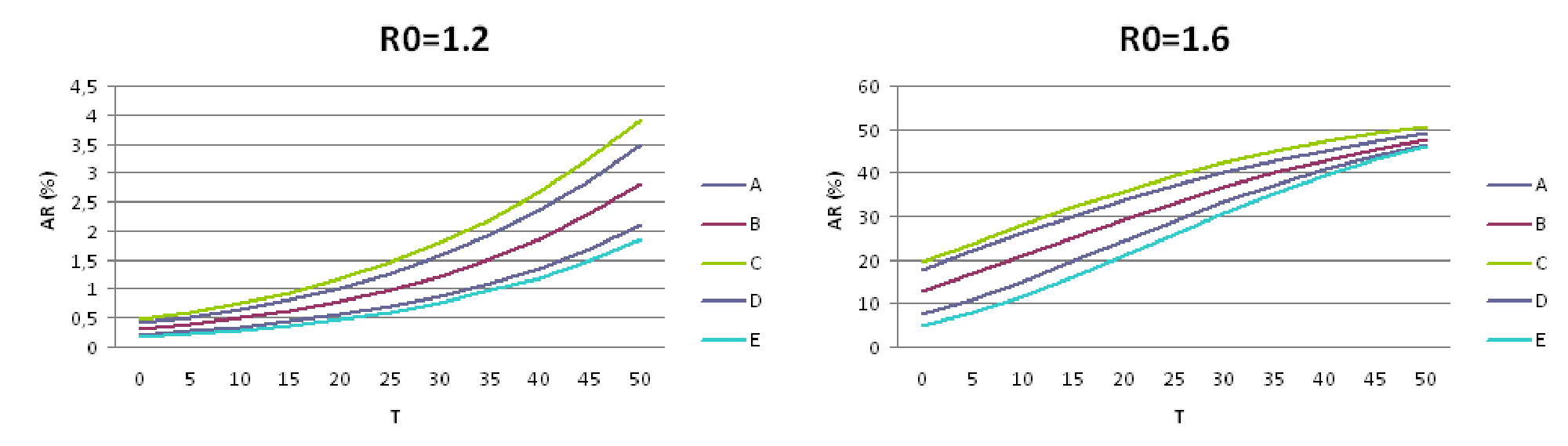}
}
\caption{Overall attack rates plotted versus the delay $T$ in start of the vaccination campaign for the cases $R_{0}=1.2$ and $R_{0}=1.6$, for the five strategies (A - Conventional, B - Uniform, C - Elderly
first, D - Children first, E - By contacts). In the absence of vaccination,
the attack rate is 22.5\% for $\ro=1.2$ and 55\% for $\ro=1.6$}
\end{figure}

{\bf Time duration to develop immunity after vaccination} 

\medskip

Ignoring the time duration $1/\mu_W$ needed to develop immunity decreased
the attack rates significantly, in some cases by 8$\%$ even for the best 
strategy. This shows the importance of incorporating this time period into
our model, otherwise the attack rates are seriously underestimated.

\medskip

{\bf Applicability of the model} 

\medskip

To illustrate the applicability of our approach, we fitted the model to the
Hungarian data of the first wave of A(H1N1)v. The epidemic curve was
reconstructed using the public reports of the National Center for Epidemiology (www.oek.hu). For the simulation, we fixed
the epidemiological parameters as in Table 1, employed publicly available vaccination data (www.jarvany.hu) and performed a grid search with respect to the basic reproduction number and the reduction of contacts during holidays to find the best fit by means of ordinary least square method. The result can be seen on Figure 8, where day 1 corresponds to September 1 and $R_{0}\approx 1.3$. The immunizations started on day 29, but with very low intensity first. The intensity increased only in a later phase of the outbreak. In Hungary, roughly 30$\%$ of the population received the vaccine in a three months period. The
decrease in cases after day 50 is due to a one week national school holiday.
Note that we included this very preliminary figure solely for the sake of illustration, to demonstrate
the applicability of our model in a real life situation, a more
comprehensive and thorough epidemiological analysis of the Hungarian data is in progress, which is beyond the scope of this study.

\begin{figure}
\centerline{
  \includegraphics[width=8cm]{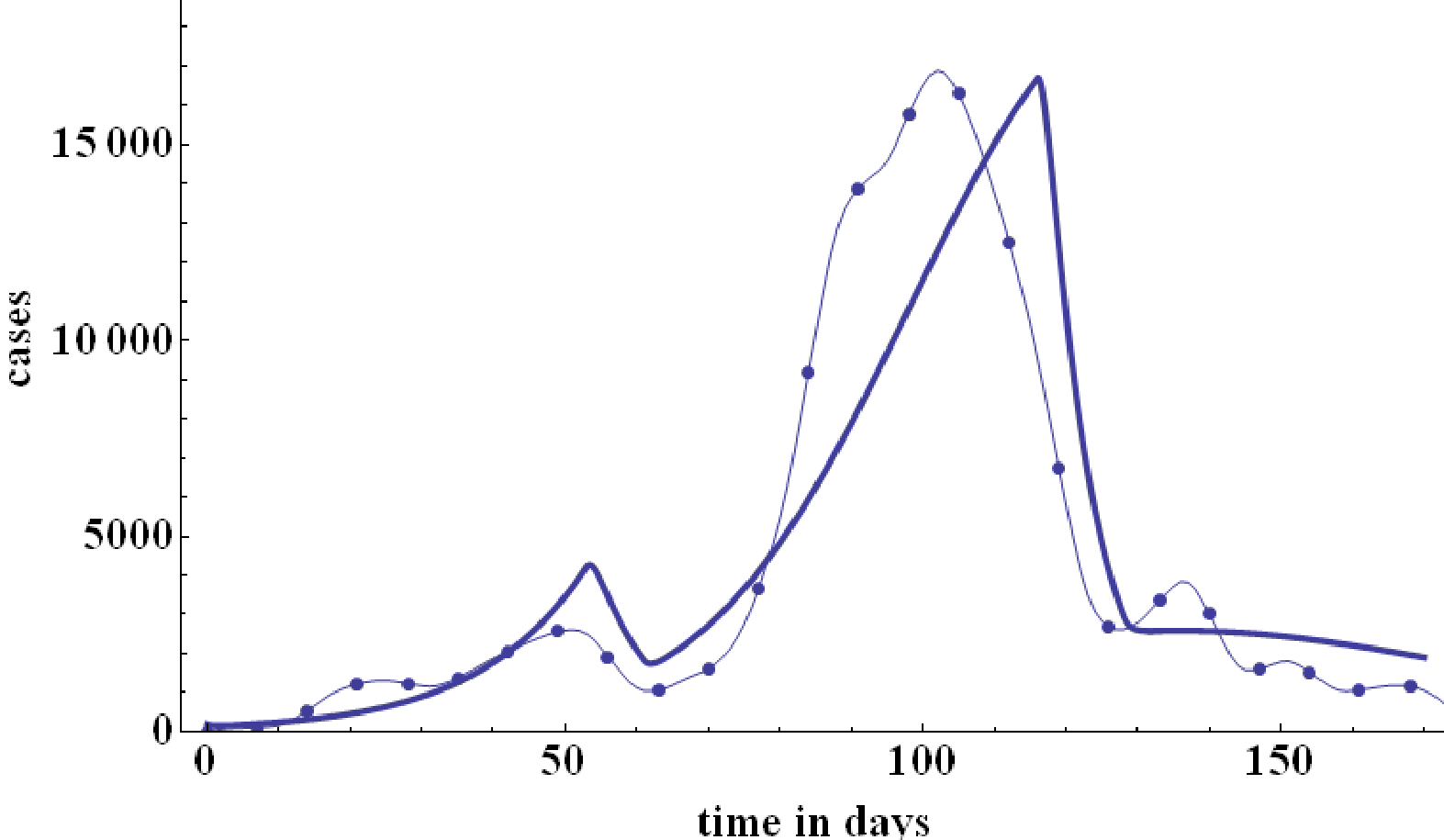}
}
\caption{Epidemic curve (thin curve with dots) in Hungary (population 10 million) and the fitted model (thick curve), taking into account the reduced
contact numbers during school holidays. Vaccinations started on day 29.}
\end{figure}

\section{Conclusions}
We have extended an established age structured mathematical population model of influenza transmission, incorporating the interplay between the vaccination campaign and the dynamics of the outbreak. The model has allowed us
to compare various age specific scheduling strategies for the overall attack rate, age specific attack rates and number of fatal cases.
We found that if 60\% coverage is targeted within each age group, the best scheduling scheme $E$ (i.e. prioritizing age groups according to the number of social contacts) can reduce the overall attack
rate by 5-10 \% and mortality by up to 30$\%$ relative to the worst strategy, depending on the delay in start of vaccination. Previous studies suggest that this can be further
optimized by allocating more vaccines to young age groups, however the scope of
our study was to examine the significance of the age specific timing in the schedule, without changing the overall coverage for the age groups. 
Our results clearly demonstrate that consideration of age specific transmission is paramount to vaccination schedule planning.

Several mathematical models have been developed to evaluate age specific vaccination strategies and the impact of timing for influenza outbreaks (Matrajt \& Longini 2010, Medlock {\it et al.} 2010, Mylius {\it et al.} 2008, Sypsa {\it et al.} 2009, Tuite {\it et al.} 2010). They mostly focused on vaccine allocation (for example in Mylius {\it et al.} 2008 and Matrajt \& Longini 2010, all vaccines are administered at once), while our main purpose was to explore the
effect of temporal order of prioritizations. Generally, recommendations depend on the progress of the epidemic. Mylius {\it et al.} 2008 and Matrajt \& Longini 2010 suggested prioritizing high risk groups if vaccine becomes available in a
later phase of the epidemic, while if available early, they suggested to protect high-transmission groups. We have consistently found that the attack rates are
the lowest when high-transmission groups are prioritized. However, this is most
important when we start the vaccination early. For a late start and higher reproduction numbers, the differences between attack rates are much smaller. Thus, if the attack rates are similar, protecting the most vulnerable first can be a better strategy resulting lower mortality. Medlock {\it et al.} 2010 was led to a similar conclusion, hence our findings are in accordance with previous results in the literature. 
However, our conclusions can be viewed as a refinement of earlier results in the sense that in our model vaccines are administered continuously, and using our methodology we can easily compare the
attack rates of various scheduling schemes for any delay in start of the campaign and any daily vaccine uptake.
We demonstrated that, besides allocation, the scheduling of age groups itself can have a huge impact on the outcome of the epidemic, especially when vaccine is available early and the
reproduction number is relatively low.

Since optimizing public health responses to this new pathogen requires difficult decisions over short timelines, a significant advantage of our approach compared to some other (stochastic, or
agent based network) models in the literature is its relative simplicity. It can be easily reproduced by other researchers and adapted to the situations of various countries where the availability of vaccine will be delayed, to identify better strategies to  mitigate the 
burden of the pandemic. Also, such strategies can be translated into
a realizable public health policy. Other real life concerns, such as to protect the most vulnerable by early vaccination of high risk groups, or 
vaccinating health care and emergency personnel, and constraints in
vaccine availability or the capacity of the health care system can be readily incorporated. Furthermore, the uncertainties involved in the parameters were assessed through a sensitivity analysis to examine whether such variation results in markedly different outcomes. The sensitivity analysis showed the
robustness of our results. However, the strategy of public health authorities must be adaptive (see Chowell {\it et al.} 2009 for a Mexican case study), especially when there is a race between the spread of the infection and the vaccination campaign, and certain parameters are clarified only during the outbreak. As the mortality pattern becomes clear as the pandemic progresses, it might be necessary to switch to a different strategy to minimize the number of fatal cases.

Finally, we wish to emphasize the utmost importance of the early start of the vaccination campaign: ten days delay may cause a significant, up to 6 \% increase in the overall attack rate.  

%\end{multicols}

\bigskip {\bf Acknowledgements:}\\
DK was supported by the Hungary-Serbia IPA Cross-border Co-operation programme, HU-SRB/0901/221/088. GR was supported by the Hungarian Research Fund grant OTKA K75517, the
Bolyai Research Scholarship of the Hungarian Academy of Sciences, and the T\'{A}MOP-4.2.2/08/1/2008-0008 program of the Hungarian National
Development Agency. The authors are grateful for the valuable comments of the
referees, and Beatrix Oroszi for helpful discussions.

\bigskip

{\bf References}

\bigskip

Alexander ME, Bowman CS, Feng Zh, Gardam M, Moghadas SM, Röst G, Wu J, Yan P 2007 Emergence of drug-resistance: implications for antiviral control of influenza pandemic P Roy Soc B - Biol Sci {\bf 274}:(1619) 1675-1684

\bigskip

Alexander ME, Moghadas SM, Röst G, Wu J 2008 A delay differential model for pandemic influenza with antiviral treatment, Bulletin of Mathematical Biology {\bf 70}(2), 382-397

\bigskip

Arino J, Brauer F, van den Driessche P, Watmough J, Wu, J 2006 Simple models for containment of a pandemic J R Soc Interface
{\bf 3}(8) 453-457

\bigskip

Baccam P, Beauchemin C, Macken CA, Hayden FG, Perelson AS. 2006 Kinetics of influenza A virus infection in humans J Virol {\bf 80}(15):7590-9.

\bigskip

Balcan D et al. 2009
Seasonal transmission potential and activity peaks of the new influenza A(H1N1): a Monte Carlo likelihood analysis based on human mobility
BMC Medicine {\bf 7}(45) doi:10.1186/1741-7015-7-45

\bigskip

Basta NE, Halloran EM, Matrajt L, Longini IM Jr. 2008 Estimating Influenza Vaccine Efficacy From Challenge and Community-based Study Data, Am J Epidemiol
{\bf 168}(12), 1343-52

\bigskip

%Chowell G, Bertozzi SM, Colchero A, Lopez-Gatell H, Alpuche-Aranda C, Hernandez %M, Miller MA 2009 Severe Respiratory Disease Concurrent with the Circulation of %H1N1  Influenza, N Eng J Med {\bf 361} 674-679	
%\bigskip

Chowell G, Ammon CE, Hengartner NW, Hyman JM 2006 Transmission dynamics of the great influenza pandemic of 1918 in Geneva, Switzerland: Assessing the effects of hypothetical interventions. J Theor Biol {\bf 241}(2), 193-204

\bigskip

Chowell G, Miller MA, Viboud C 2007 Seasonal influenza in the United States, France and Australia: Transmission and prospects for control. Epidemiol Infect {\bf 136}(06) 852--864

\bigskip

Chowell G, Viboud C, Wang X, Bertozzi S, Miller M, 2009, Adaptive vaccination strategies to mitigate pandemic influenza - Mexico as a case study,
PLoS ONE {\bf 4}(12): e8164 

\bigskip

CDC 2009, Key Facts About Seasonal Flu Vaccine, 

http://www.cdc.gov/Flu/protect/keyfacts.htm 

\bigskip

CDC 2008, Prevention and Control of Influenza: Recommendations of the Advisory Committee on Immunization Practices (ACIP),  July 17, 2008 / 57 (Early Release);1-60

\bigskip

Diekmann O, Heesterbeek JAP, Roberts MG, The construction of next-generation matrices for compartmental epidemic models, J R Soc Interface {\bf 7}(47) 873-885 

\bigskip

Donaldson LJ, Rutter PD, Ellis BM, Greaves FEC, Mytton OT, Pebody RG, Yardley IE, Mortality from pandemic A/H1N1 2009 influenza in England: public health surveillance study, British Medical Journal,{ \bf 339}, {b5213}, 2009. 

\bigskip

European CDC Risk Assessment 2009, Pandemic H1N1, Version 6, 6 November 2009 

\bigskip

Eurostat 2006, Population Statistics 2006 edition, Official Publications of the European Communities (Luxembourg)

\bigskip

Falus F, Oroszi B, 2009, National Chief Medical Officer's Office, Hungary, personal
communication

\bigskip

Ferguson NM, Cummings DA, Fraser C, Cajka JC, Cooley PC:  Strategies for mitigating an influenza pandemic. 2006 Nature {\bf 442}(7101):448-452.

\bigskip
Fleming DM, Elliot AJ 2006 Estimating the risk population in relation to influenza vaccination policy, Vaccine {\bf 24}(20), {4378--4385},

%Fisman, D. N., Savage, R., Gubbay, J., Achonu, C., Akwar, H., Farrell, D. J., %Crowcroft, N. S., Jackson, P. 2009, Older Age and a Reduced Likelihood of 2009 %H1N1  Virus Infection. N Eng J Med {\bf 361} 2000-2001

\bigskip

Gojovic MZ, Sander B, Fisman D, Krahn MD, Bauch CT 2009 
Modelling mitigation strategies for pandemic (H1N1) 2009, Can. Med. Assoc. J., {\bf 181}(10):673 
10.1503/cmaj.091641

\bigskip

Goldstein E, Apolloni A, Lewis B, Miller JC, Macauley M, Eubank S,
Lipsitch M, Wallinga J 2009 Distribution of vaccine/antivirals and the least spread line in a stratified population, J R Soc Interface {\bf 7}(46)  755-764 

\bigskip

Greenberg ME, Lai MH, Hartel GF, Wichems CH, Gittleson C, Bennet J, Dawson G, Hu W, Leggio C, Washington D, Basser RL 2009
Response to a Monovalent Influenza A (H1N1) 2009 Vaccine, N Eng J Med {\bf 361}(25):2405-2413 

\bigskip

Hancock K, Veguilla V, Lu X, et al. Cross-reactive antibody responses to the 2009 pandemic H1N1  influenza virus. N Engl J Med {\bf 361}(20):1945-52.

\bigskip

Heffernan JM, Smith RJ, Wahl LM 2005 Perspectives on the basic reproductive ratio. J R Soc Interface {\bf 2}(4):281-93

\bigskip

Johansen K, Nicoll A, Ciancio BC, Kramarz P 2009 Pandemic influenza A(A(H1N1)v) 2009 vaccines in the European Union. Euro Surveill {\bf 14}(41): pii=19361

\bigskip

Longini IM Jr, Halloran ME 2005 Strategy for distribution of influenza vaccine to high-risk groups and children, Am J Epidemiol {\bf 161}(4), 303

\bigskip

Matrajt L, Longini IM 2010 Optimizing Vaccine Allocation at Different Points in Time During an Epidemic, UW Biostatistics Working Paper Series 363

\bigskip

Medlock J, Galvani AP 2009 Optimizing influenza vaccine distribution, Science {\bf 325}(5948), 1705 - 1708

\bigskip

Medlock J, Meyers LA, Galvani AP 2010 Optimizing Allocation for a Delayed Influenza Vaccine Campaign, PLOS Curr Influenza, RRN1134

\bigskip

Merler S, Ajelli M, Rizzo C 2009 Age-prioritized use of antivirals during an influenza pandemic
{BMC Infectious Diseases} {\bf 9}:{117}

\bigskip

Miller E, Hoschler K, Hardelid P, Stanford E, Andrews N, Zambon M 2010
{Incidence of 2009 pandemic influenza A H1N1 infection in England: a cross-sectional serological study}, {The Lancet} {\bf 375}(9720):1100-1108 

\bigskip

Moghadas SM, Bowman CS, Röst G, Wu J 2008 Population-wide emergence of antiviral resistance during pandemic influenza, PLOS ONE {\bf 3}(3) e1839

\bigskip

Moghadas SM, Bowman CS, Röst G, Fisman D, Wu J 2009 Post-exposure prophylaxis during pandemic outbreaks, BMC Medicine, {\bf 3}(73)

\bigskip

Moghadas S, Day T, Bauch C, Driedger SM, Brauer F, Greer AL, Yan P, Wu J, Pizzi N, Fisman DN 2009 Modelling an influenza pandemic: a guide for the perplexed, Can. Med. Assoc. J., August 4, 2009; {\bf 181} (3-4), 171-3

\bigskip

Mossong J, Hens N, Nit M et al. 2008 Social contacts and mixing patterns relevant to the spread of infectious diseases, PLOS Medicine, {\bf 5}(3):e74

\bigskip

Mylius SD, Hagenaars TJ, Lugnér AK, Wallinga J 2008 Optimal Allocation of Pandemic Influenza Vaccine Depends on Age, Risk and Timing, Vaccine, {\bf 26}, 3742-3749

\bigskip

Nichol KL, Efficacy/clinical effectiveness of inactivated influenza virus vaccines in adults 1998, In: Nicholson KG, Webster RG, Hay AJ, editors. Textbook of influenza. Malden, MA: Blackwell Science Ltd., 1998:358 -72

\bigskip

Nishiura H, Iwata K 2009 A simple mathematical approach to deciding the dosage of vaccine against pandemic H1N1  influenza. Euro Surveill {\bf 14}(45)
pii=19396

\bigskip

Presanis AM,  De Angelis D,  Hagy A,  Reed C,  Riley S,  et al. 2009 The Severity of Pandemic H1N1 Influenza in the United States, from April to July 2009: A Bayesian Analysis. PLoS Med {\bf 6}(12): e1000207  

\bigskip

Robinson R, A(H1N1)v  vaccine products and production, ACIP July 2009 meeting, www.cdc.gov/vaccines/recs/acip/downloads/mtg-slides-jul09-flu/05-Flu-Robinson.pdf

\bigskip

Serfling RE, Sherman IL, Houseworth WJ 1967 Excess pneumonia-influenza mortality by age and sex in three major influenza A2 epidemics, United States, 1957-58, 1960 and 1963. Am J Epidemiol {\bf 86}(2):433-41.

\bigskip

Sypsa V, Pavlopoulou I, Hatzakis A 2009 Use of an Inactivated Vaccine in Mitigating Pandemic Influenza A(H1N1) Spread: A Modelling Study to Assess the Impact of Vaccination Timing and Prioritisation Strategies, Euro Surveill, 2009:14(41):pii=19356

\bigskip

Tuite AR, Fisman DN, Kwong JC, Greer AL 2010 Optimal Pandemic Influenza Vaccine Allocation Strategies for the Canadian Population, PLOS ONE, {\bf 5}(5):e10520.

\bigskip

Tuite AR, Greer AL, Whelan M, Winter AL, Lee B, Yan P, Wu J, Moghadas SM, Buckeridge D, Pourbohloul B, Fisman DN 2009 Estimated epidemiologic parameters and morbidity associated with pandemic H1N1  influenza, Can Med Assoc J {\bf 182}(2):131

\bigskip

Vaillant L, La Ruche G, Barboza P, for the epidemic intelligence team at InVS. Epidemiology of fatal cases associated with pandemic H1N1 influenza 2009. Euro Surveill {\bf 14}(33):pii=19309

\bigskip

Vajo Z, Wood J, Kosa, Szilvasy I, Paragh Gy, Pauliny Zs, Bartha K, Visontay I, Kis A, Jankovics I 2009 
A single-dose influenza A (H5N1) vaccine safe and immunogenic in adult and elderly patients - an approach to pandemic vaccine development
 J Virol {\bf 84}(3):1237-1242 

\bigskip

Yang Y, Sugimoto JD, Halloran ME, Basta NE, Chao DL, Matrajt L, Potter G, Kenah E, Longini Jr IM 2009 The transmissibility and control of pandemic influenza A (H1N1) virus, 2009 {Science}, {\bf 326}(5953):729-33

\end{document}